\documentclass[a4paper,twocolumn,english,prl]{revtex4}
\usepackage[T1]{fontenc}
\usepackage[latin1]{inputenc}
\usepackage{graphicx}

\makeatletter

\date{\today}
\usepackage{ae,aecompl}

\usepackage{bm}

\usepackage{babel}
\makeatother
\begin{document}

\title{Quantum phase-space simulations of fermions and bosons}

\author{P. D. Drummond and J. F. Corney}

\address{ARC Centre of Excellence for Quantum-Atom Optics, University of Queensland,
Brisbane 4072, Queensland, Australia.}

\begin{abstract}
We introduce a unified Gaussian quantum operator representation for
fermions and bosons. The representation extends existing phase-space
methods to Fermi systems as well as the important case of Fermi-Bose
mixtures. It enables simulations of the dynamics and thermal equilibrium
states of many-body quantum systems from first principles. As an example,
we numerically calculate finite-temperature correlation functions
for the Fermi Hubbard model, with no evidence of the Fermi sign problem.
\end{abstract}

\maketitle
Calculating the quantum many-body physics of interacting Fermi systems
is one of the great challenges in modern theoretical physics. In even
the simplest cases, first-principles calculations are made difficult
by the complexity of the fermionic wave-function, manifest notoriously
in the Fermi sign problem. In previous quantum Monte Carlo (QMC) techniques,
the sign problem appears as trajectories with negative weights, which
contribute to a large sampling error\cite{Ceperley99}, together with
large, computationally intensive determinants.

Fermion complexity issues appear in physical problems at all energy
scales, from high-energy lattice QCD to the emerging area of ultra-cold
atomic physics. Recent pioneering experiments in ultra-cold Fermi
gases are capable of investigating fermion many-body physics in regimes
of unprecedented experimental simplicity. This situation implies a
substantial opportunity to develop and test novel first-principles
theoretical methods for the investigation of correlations and dynamical
effects.

Here we present a phase-space method for simulating many-body
boson\cite{Gauss:Bosons,PRL:Bosons} and  fermion\cite{Gauss:Fermi-Bose,Gauss:Fermions} systems, based on a Gaussian 
operator expansion.

The method allows the treatment of dynamical and static problems at
finite temperature. The expansion in the fermionic case represents
\emph{pairs} of Fermi operators. Since the pairs obey commutation
relations, there are no anti-commutators causing sign problems, and
no large determinant calculations as in some previous approaches.

The method is illustrated using the finite temperature Hubbard model,
which is a well-known theory in condensed matter physics and high
$T_{c}$ superconductors. The cases chosen have an acute sign problem
using conventional QMC. The results are directly applicable to 
feasible experiments on ultra-cold
fermions in an optical lattice\cite{Fermilattice}. 

Like path-integral QMC, phase-space methods sample the many-body quantum density operator $\widehat\rho$.  But rather than expressing the density operator in a position representation, one expands it in terms of an overcomplete basis of operators:
\begin{equation}
\widehat{\rho}(t)=\int P(\overrightarrow{\lambda},t)\widehat{\Lambda}(\overrightarrow{\lambda})d\overrightarrow{\lambda}\,\,,\label{eq:general-expansion}\end{equation}
where $P(\overrightarrow{\lambda},t)$ is a positive probability distribution,
$\widehat{\Lambda}$ is an overcomplete operator basis for the class of density
matrices being considered, and $d\overrightarrow{\lambda}$ is the
integration measure for the generalized phase-space coordinate $\overrightarrow{\lambda}$.   It is the overcompleteness of the basis which allows a positive representation of any physical density matrix in terms of Gaussian operators.

We define the operator basis $\widehat{\Lambda}\equiv\Omega\widehat{\Lambda}_{+}\widehat{\Lambda}_{-}\,\,$
to be the product of Gaussian forms of bosonic ( $\widehat{\Lambda}_{+}$)
and fermionic ( $\widehat{\Lambda}_{-}$) creation and annihilation
operators, over $M_{\pm}$ modes respectively. Here $\Omega$ is an
additional weighting factor. If $\widehat{\mathbf{a}}$ is a row vector
of $M_{\pm}$ annihilation operators, and $\widehat{\mathbf{a}}^{\dagger}$
the corresponding column vector of creation operators, their commutation
relations are: $[\widehat{a}_{k},\widehat{a}_{j}^{\dagger}]_{\mp}=\delta_{kj}\,\,.$
We use $\mp$ to indicate bosons (upper sign) or fermions (lower sign). 
For brevity, we restrict the present discussion to number-conserving systems.  The most general Gaussian operator is then a generalized thermal density operator with a complex covariance:

\begin{eqnarray}
\widehat{\Lambda}_{\pm}(\mathbf{n})  &=&  \left|\mathbf{I}\pm\mathbf{n}\right|^{\mp1}:\exp\left[\widehat{\mathbf{a}}\left(\left\{ 2\mathbf{I}\right\} -\left[\mathbf{I}\pm\mathbf{n}\right]^{-1}\right)\widehat{\mathbf{a}}^{\dagger}\right]:\,\,\nonumber \\ 
\label{eq:Gaussbasis}\end{eqnarray}
 where the additional $\left\{ \mathbf{I}\right\} $ in the exponent
is bracketed to indicated that it only appears in the fermionic case.
Normal ordering is defined as usual for Bose and Fermi systems, for example, $:\widehat{a}_{j}\widehat{a}_{i}^{\dagger}:\,=\pm\widehat{a}_{i}^{\dagger}\widehat{a}_{j}=\pm\widehat{n}_{ij}\,\,$.
The $M_{\pm}\times M_{\pm}$ matrix $\mathbf{n}$ corresponds to a
generalized thermal covariance. 

Using these Gaussian operators in the density operator expansion of Eq.\ (1), one finds that operator expectation values become
weighted moments of the distribution $P$, denoted as $\left\langle ..\right\rangle _{P}$.  Thus the first- and second-order number correlations are
\begin{eqnarray}
&&\left\langle \widehat{\mathbf{n}}\right\rangle =\int\Omega\mathbf{n}
P(\overrightarrow{\lambda},t)d\overrightarrow{\lambda}\,=
\left\langle \mathbf{n}\right\rangle _{P}\,\, ,\nonumber \\
&&\left\langle \widehat{a}^\dagger_i\widehat{a}^\dagger_j\widehat{a}_j\widehat{a}_i \right\rangle=\left\langle n_{ii}n_{jj}\right\rangle_P \pm \left\langle n_{ij}n_{ji}\right\rangle_P\, . 
\end{eqnarray}

As an illustration of the use of the unified representation, consider
the canonical distribution of a Bose or Fermi field. The thermal state
at temperature $T=1/k_{B}\tau$ can be cast into an imaginary time
integro-differential equation: \begin{equation}
\frac{d\widehat{\rho}}{d\tau}=
-\frac{1}{2}\int P(\overrightarrow{\lambda},t)[\widehat{H}\,,
\widehat{\Lambda}(\overrightarrow{\lambda})]_{+}d\overrightarrow{\lambda}\,\,.
\label{eq:prop}
\end{equation}
To solve this, we first use identities derived in \cite{Gauss:Bosons,Gauss:Fermions} that describe the action of operators on the density operator as derivatives on elements of the Gaussian
basis. After integrating Eq (\ref{eq:prop}) by parts, we arrive at the following mappings:
\begin{eqnarray}
\widehat{\mathbf{n}}\widehat{\rho} & \rightarrow & 
\mathbf{n}P-\left\{\frac{\partial }{\partial\mathbf{n}}\left(\mathbf{I}\pm\mathbf{n}\right)^T\right\}^T\mathbf{n}P\nonumber \\
\widehat{\rho}\widehat{\mathbf{n}} & \rightarrow & \mathbf{n}P-\left\{\frac{\partial }{\partial\mathbf{n}}\mathbf{n}^T\right\}^T\left(\mathbf{I}\pm\mathbf{n}\right)P \,\,.\end{eqnarray}
The matrix derivative is here defined as $(\partial/\partial\mathbf{n})_{ij}=\partial/\partial n_{ij}\,\,$.
In the free-field case of $\widehat{H}=\sum_{k}\omega_{k}\widehat{n}_{k}$,
we arrive at a first-order Fokker-Planck equation for the distribution
function $P$. This leads to deterministic equations for the mode
occupations of form $\partial n_{k}/\partial\tau=-\omega_{k}n_{k}\left(1\pm n_{k}\right)\,\,,$
which can be integrated to get the well-known Bose-Einstein (Fermi-Dirac)
distribution: 
\begin{equation}
n_{k}=\frac{1}{\exp(\omega_{k}\tau)\mp1} \,\, .
\end{equation}

For systems of {\emph interacting} particles, the unified representation gives
nonlinear, stochastic phase-space equations, which must be solved
numerically. Due to the non-uniqueness of the expansion, a careful
choice of identities is mandatory\cite{Gauge} to
keep the nonlinear equations stable. We term this choice a stochastic
gauge in analogy to the gauge fields in QED, since it results in freely chosen
fields in the resulting stochastic equations. The choice in the fermionic case 
is especially large, since the fermionic anti-commutation relations result in
a non-unique algebraic form of the Hamiltonian.

As an example, consider the Hubbard model, which is the simplest nontrivial model for strongly interacting electrons.  It is an important system in condensed matter physics, with
relevance to the theory of high-temperature superconductors\cite{Linden92}.
The Hamiltonian is:
\begin{eqnarray}
H(\widehat{\mathbf{n}}_{1},\widehat{\mathbf{n}}_{-1})&=&
-\sum_{ij,\sigma}\left(t_{ij}+\mu\delta_{ij}\right)\widehat{n}_{ij,\sigma}
\nonumber \\
&&-|U|\sum_{j}:
\left(\widehat{n}_{jj,1}-s\widehat{n}_{jj,-1}\right)^{2}:/2
\label{eq:Hubbard}
\end{eqnarray}
 where 
$\widehat{n}_{ij,\sigma}=\widehat{a}_{i,\sigma}^{\dagger}\widehat{a}_{j,\sigma}=
 \left\{ \widehat{\mathbf{n}}_{\sigma}\right\} _{ij}$.
The coupling $t_{ij}=t$ if the $i,j$ correspond to nearest neighbour
sites and is otherwise $0$. The index $\sigma$ denotes spin ($\pm1$)
and the indices $i,j$ label lattice location, and $s=U/|U|=\pm1$.
Traditional QMC methods for this problem have large sign problems
in the repulsive case ($s=1$)\cite{Santos03,FettesMorgenstern00}. 

Because of the way we have written the interaction term in Eq.\ (6) (which constitutes a kind of fermionic gauge choice), the Hubbard model maps to a set of \emph{}real Stratonovich stochastic
equations:\begin{eqnarray}
\frac{d\mathbf{n}_{\sigma}}{d\tau} & = & 
\frac{1}{2}\left\{ \left(\mathbf{I}-\mathbf{n}_{\sigma}\right)\bm T_{\sigma}^{(1)}\!\mathbf{n}_{\sigma}+
\mathbf{n}_{\sigma}\bm T_{\sigma}^{(2)}\!\left(\mathbf{I}-\mathbf{n}_{\sigma}\right)\right\} .\nonumber \\
\end{eqnarray}
Here we have introduced the stochastic propagation matrix:
\begin{eqnarray}
T_{ij,\sigma}^{(r)}&=&t_{ij}+\delta_{ij}\left\{\mu-
\sigma^{(s+1)/2}\xi_{j}^{(r)}\right\}
 \nonumber\\
&&-
|U|\delta_{ij}\left\{ sn_{jj-\sigma}-n_{jj\sigma}+\frac{1}{2}\right\}
 .\end{eqnarray}
 The real Gaussian noise $\xi_{j}^{(r)}(\tau)$ is defined by the
correlations 
\begin{equation}\left\langle \xi_{j}^{(r)}(\tau)\,\xi_{j'}^{(r')}(\tau')\right\rangle =2|U|\delta(\tau-\tau')\delta_{jj'}\delta_{rr'}\,\,.
\end{equation}

The weights for each trajectory evolve as physically expected for
energy-weighted averages, with $d\Omega/d\tau=-\Omega H(\mathbf{n}_{1},\mathbf{n}_{-1})$.
Because the equations for the phase-space variables $n_{ij,\sigma}$
are all real, the weights of all trajectories will remain positive.  Thus the traditional manifestation of the sign problem is avoided, as there is no `deterioration of the sign' from averaging over positive and negative weights.  Furthermore, the mapping to real phase-space produces a stable set of equations, and thus there is no need to invoke additional gauge choices.

There is, however, the issue of spreading weights, which becomes serious for large lattice sizes and long simulations times.   Physical quantities are weighted averages: \begin{equation}
\left\langle A(\tau) \right\rangle_P = \sum_{p=1}^{N_p}\Omega^{(p)}(\tau)A^{(p)}(\tau)/\sum_{p=1}^{N_p}\Omega^{(p)}(\tau),
\end{equation}
where $N_p$ is the total number of paths in the sample.  A large spread in the weights makes a straightforward  average very inefficient, as most paths in the ensemble may end up contributing very little to the final result.  To increase efficiency, we instead use a simple branching algorithm adapted from Green's function Monte Carlo methods\cite{TrivediCeperley90}, in which low-weights paths are deleted and high-weight paths are cloned, according to the rate:
\begin{eqnarray}
m^{(p)} & = & {\rm Integer}\left[\xi+\Omega^{(p)}/\overline{\Omega}\right],
\end{eqnarray} 
where $\xi$ is a random variable uniformly distributed on $\left[0,1\right]$ and where $\overline{\Omega}$ is a reference weight, which is adapted to keep the number of paths $N_p$ under control.  At branching, the weights are equalised and thereafter the clones evolve independently with spreading weights.  To avoid biasing, the branching must occur sufficiently often to limit the diversity of weights at the branching times. For the results presented here, the branching algorithm is sufficient to control sampling error - other situations may require the use of more sophisticated importance sampling methods\cite{Linden92}.

The stochastic phase-space equations are simulated by a robust semi-implicit algorithm\cite{DrummondMortimer91}, with an adaptive stepsize to overcome stiffness.  Unlike Projector QMC methods, the Gaussian phase-space method can calculate any correlation function, at any temperature. Unlike Path Integral QMC, a single run generates results for a range of temperatures: longer simulation times correspond to lower temperatures.  Strictly speaking, zero-temperature results are obtained only in the limit of long simulation times.  In practice, however, one only has to run the simulation until the relevant correlation functions have plateaued.

Precision is of course limited by sampling error, but this can be reduced by several means.  For example, one can a) include more trajectories in the sample, b) employ a more sophisticated branching/importance sampling technique to reduce the spread in weights, and c) make a better `stochastic gauge' choice to obtain phase-space equations with smaller sampling error for the correlations to be calculated.

Typical results for a $16\times16$ lattice are shown in Figs (1) and (2), which plot the energy $E$ and second-order correlation function $g_2$, respectively, for different chemical potentials.  The estimation of sampling error shown in the figures assumes independent samples, for simplicity.  While this is liable to underestimate the error, especially for $g_2$ where there is also spatial averaging, it does indicate the approximate dependence on temperature.  In particular, the sampling error remains well-controlled throughout the simulation, even away from half filling, where there is known to be a sign problem.  A more detailed sampling error analysis will be given elsewhere.

\begin{figure}
\begin{center}\includegraphics[%
  width=7cm]{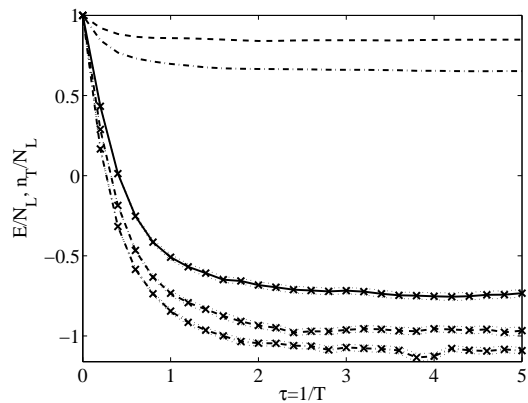}\end{center}

\caption{\label{cap:Total-energy}Total energy $E$ per site versus inverse
temperature $\tau$ for a $16\times16$ 2D lattice for chemical potentials
$\mu=2$ (solid), $\mu=1$ (dashed) and $\mu=0$ (dot-dashed). Curves
without crosses give the number of particles per site for $\mu=1$
(dashed) and $\mu=0$ (dot-dashed). $U=4$, $t=1$, and 100 paths initially.
Dotted curves give an approximate sampling error.}
\end{figure}

\begin{figure}
\begin{center}\includegraphics[%
  width=7cm]{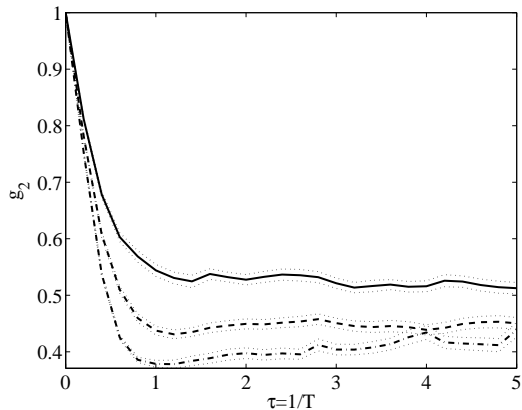}\end{center}

\caption{\label{cap:g2}Second-order correlation function $g_2\equiv \left\langle \sum_{j}\widehat{n}_{jj,1}\widehat{n}_{jj,-1}\right\rangle/\left\langle \sum_{j}\widehat{n}_{jj,1}\right\rangle\left\langle \sum_{j}\widehat{n}_{jj,-1}\right\rangle$ versus inverse
temperature $\tau$ for a $16\times16$ 2D lattice for chemical potentials
$\mu=2$ (solid), $\mu=1$ (dashed) and $\mu=0$ (dot-dashed). $U=4$, $t=1$, and 100 paths initially.
Dotted curves give an approximate sampling error.}
\end{figure}

In conclusion, we have  presented a unified operator representation that is able to represent arbitrary physical states of bosons and fermions.  By use of this representation, non-interacting systems can be mapped to deterministic phase-space equations, whereas systems with two-body interactions can be simulated by use of stochastic sampling methods, provided a suitable gauge is chosen to eliminate any boundary terms.  For the example of the Hubbard model, we show how the thermal equilibrium problem can be mapped to a set of real, stable phase-space equations with positive weights.  Bosonic problems can be solved in similar manner, and the method can also be used to simulate dynamics (although typically with an increased sampling error).  Thus the one, unified method can solve
both fermionic and bosonic problems, which  makes it well suited to simulating Bose-Fermi mixtures, and to studying the BEC/BCS crossover.\bibliographystyle{elsart-num}
\bibliography{CCP2004DC4}

\end{document}